\documentclass[12pt]{article}
\usepackage{epsfig}

\def\beq{\begin{equation}}
\def\eeq#1{\label{#1}\end{equation}}
\def\eeqn{\end{equation}}

\def\beqa{\begin{eqnarray}}
\def\eeqa#1{\label{#1}\end{eqnarray}}
\def\eeqan{\end{eqnarray}}

\let\bar=\overbar

\def\Dslash{\not{\hbox{\kern-4pt $D$}}}
\def\dslash{\not{\hbox{\kern-2pt $\del$}}}

\def\msb{{\bar{\ssstyle M \kern -1pt S}}}

\def\beq{\begin{equation}}
\def\eeq{\end{equation}}
\def\beqa{\begin{eqnarray}}
\def\eeqa{\end{eqnarray}}
 
\def\Title#1{\begin{center} {\Large {\bf #1} } \end{center}}

\begin{document}

\Title{Charged Higgs production with a $W$ boson via $b$-quark annihilation}

\bigskip\bigskip

\begin{raggedright}  

{\it Nikolaos Kidonakis\footnote{Talk presented at the APS Division of Particles
 and Fields Meeting (DPF 2017), July 31-August 4, 2017, Fermilab. C170731}\\
Department of Physics\\
Kennesaw State University\\
Kennesaw, GA 30144, USA}
\bigskip\bigskip
\end{raggedright}

\section{Introduction}

I present theoretical calculations for charged Higgs production in association with a $W$ boson via bottom quark annihilation. I discuss higher-order radiative corrections from collinear and soft gluon emission and show that they are important. I present theoretical results for total cross sections, and for transverse-momentum and rapidity distributions of the charged Higgs boson at LHC energies.

\section{Higher-order corrections}

Charged Higgs bosons appear in new physics models such as 2-Higgs doublet 
models, and they have been searched for extensively at the LHC.
In this paper we consider charged Higgs production in association with a $W$ boson via $b$-quark annihilation, $b{\bar b} \rightarrow H^- W^+$ \cite{HWcorr,bbHW}. As is also known from related processes, such as $tH^-$ production \cite{NKtH}, higher-order corrections are significant. 
Given the very massive final state in  $H^- W^+$ production, soft and collinear gluon corrections are important.

\begin{figure}[htb]
\begin{center}
\epsfig{file=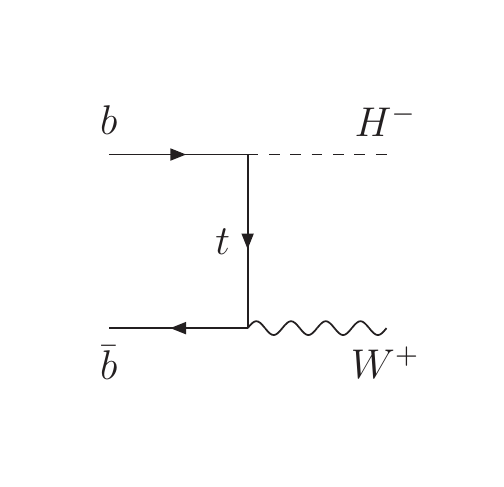,height=2.3in}
\hspace{3mm}
\epsfig{file=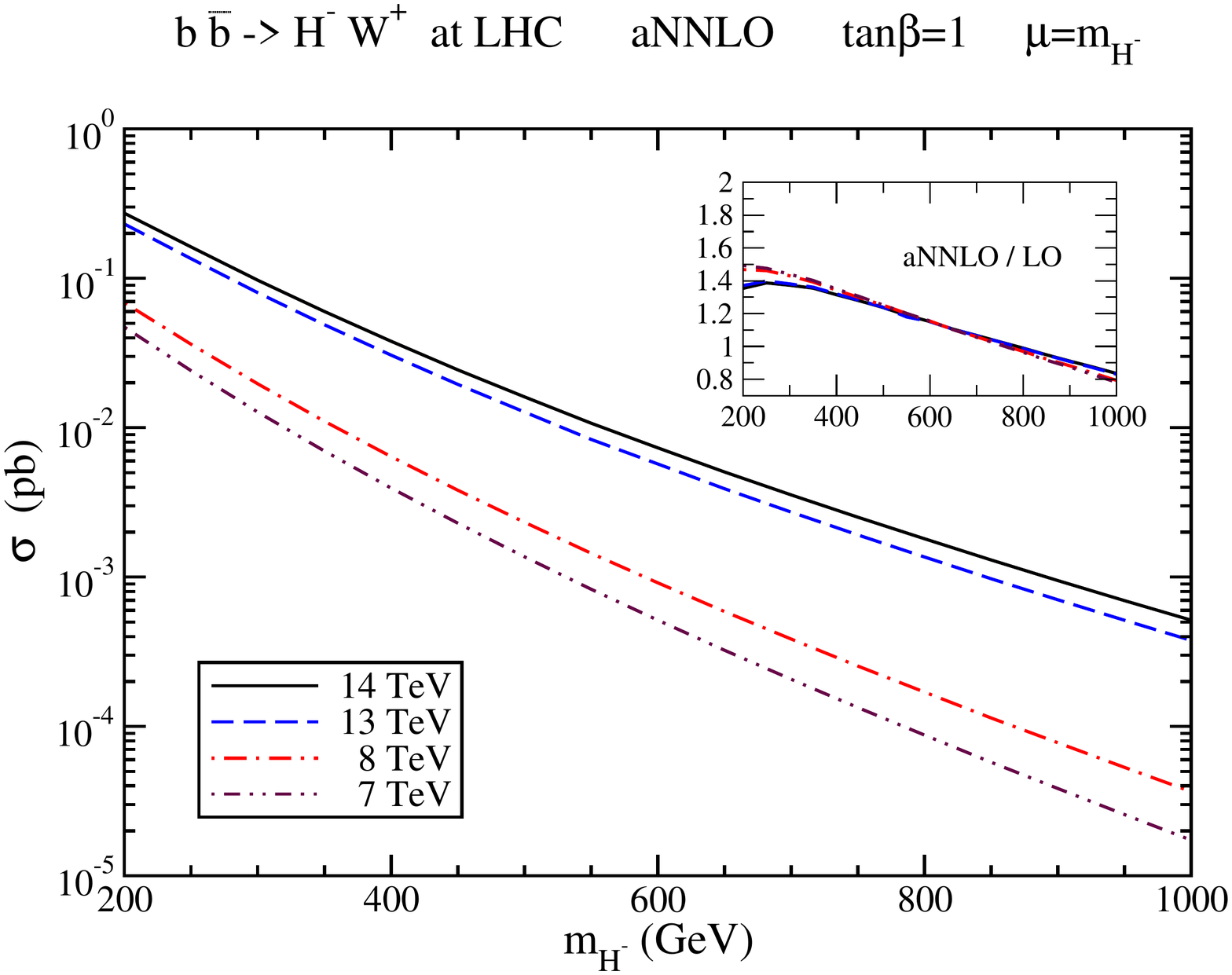,height=2.5in}
\caption{(Left) Leading-order diagram for $b{\bar b} \rightarrow H^-W^+$. 
(Right) Total cross section at aNNLO for $H^-W^+$ production at LHC energies.}
\label{bbhw}
\end{center}
\end{figure}

For the partonic process
$b(p_1)\, + \, {\bar b}\, (p_2) \rightarrow H^-(p_3)\, + W^+(p_4)$
we define $s=(p_1+p_2)^2$, $t=(p_1-p_3)^2$, $u=(p_2-p_3)^2$.  
In the left plot of Fig. \ref{bbhw} we show the lowest-order diagram for 
this process. We also define the threshold variable $s_4=s+t+u-m_H^2-m_W^2$, 
which vanishes at partonic threshold.
Soft-gluon corrections appear in the perturbative cross section as
$[\ln^k(s_4/m_H^2)/s_4]_+$. Terms of the form 
$\ln^k(s_4/m_H^2)$, of purely collinear origin, also appear.
We take moments of the partonic cross section with moment variable $N$,
${\hat \sigma}(N)=\int (ds_4/s) \; e^{-N s_4/s} \; {\hat \sigma}(s_4)$, 
and then write the cross section in $4-\epsilon$ dimensions in factorized form 
\cite{bbHW},
\beq
{\hat \sigma}^{b{\bar b} \rightarrow H^- W^+}(N,\epsilon)= 
\left( \prod_{i=b,{\bar b}} J_i\left (N,\mu,\epsilon \right) \right)
H^{b{\bar b} \rightarrow H^- W^+} \left(\alpha_s(\mu)\right) 
S^{b{\bar b} \rightarrow H^- W^+} 
\left(\frac{m_H}{N \mu},\alpha_s(\mu) \right) , 
\eeq
where $\mu$ is the scale, $\alpha_s$ is the strong coupling, $J_i$ are jet functions, $H^{b{\bar b}\rightarrow H^-W^+}$ is the hard function, and $S^{b{\bar b}\rightarrow H^-W^+}$ is the soft function.

The soft function $S^{b{\bar b}\rightarrow H^-W^+}$ obeys the renormalization group equation
\beq
\left(\mu \frac{\partial}{\partial \mu}
+\beta(g_s, \epsilon)\frac{\partial}{\partial g_s}\right)\, 
S^{b{\bar b} \rightarrow H^-W^+}
=- 2 \, S^{b{\bar b} \rightarrow H^-W^+} \, \Gamma_S^{b{\bar b} \rightarrow H^-W^+}  
\eeq
where $\Gamma_S^{b{\bar b}\rightarrow H^-W^+}$ is the soft anomalous dimension.

From the evolution of the soft and jet functions we write the resummed cross section as \cite{bbHW}
\beqa
{\hat{\sigma}}_{\rm res}^{b{\bar b} \rightarrow H^- W^+}(N) &=&   
\exp\left[\sum_{i=b,{\bar b}} E_i(N_i)\right]
H^{b{\bar b} \rightarrow H^- W^+} \left(\alpha_s(\sqrt{s})\right) 
\nonumber \\ && \hspace{-15mm}\times \; 
S^{b{\bar b} \rightarrow H^- W^+}\left(\alpha_s(\sqrt{s}/{\tilde N'})\right) \; 
\exp \left[2\int_{\sqrt{s}}^{{\sqrt{s}}/{\tilde N'}} 
\frac{d\mu}{\mu}\; \Gamma_S^{b{\bar b} \rightarrow H^- W^+}
\left(\alpha_s(\mu)\right)\right] \, .
\eeqa

We use the resummed cross section in moment space as a generator of fixed-order expansions in momentum space through NNLO \cite{NNLO}.
The approximate NNLO (aNNLO) collinear and soft-gluon corrections are 
\cite{bbHW}
\beq
\frac{d^2{\hat{\sigma}}_{\rm aNNLO}^{(2) \, b{\bar b} \rightarrow H^- W^+}}{dt \, du}= 
F_{LO}^{b{\bar b} \rightarrow H^- W^+} \frac{\alpha_s^2}{\pi^2}
\left\{-C_3^{(2)} \frac{1}{m_H^2} \ln^3\left(\frac{s_4}{m_H^2}\right) 
+\sum_{k=0}^3 C_k^{(2)} \left[\frac{\ln^k(s_4/m_H^2)}{s_4}\right]_+ \right\} \, , 
\eeq
where $F_{LO}^{b{\bar b} \rightarrow H^- W^+}$ is the lowest-order term, 
and the coefficients of the two highest powers of the logarithms 
in the soft terms are 
$C_3^{(2)}=8 C_F^2$ and
\beq
C_2^{(2)}=-12 C_F^2\left(\ln\left(\frac{(t-m_W^2)(u-m_W^2)}{m_H^4}\right)
+\ln\left(\frac{\mu_F^2}{s}\right)\right)  
-\frac{11}{3} C_F C_A +\frac{2}{3} C_F n_f \, .
\eeq
Expressions for the coefficients of the lowest two powers of the logarithms in the soft terms can be found in \cite{bbHW}. Regarding the purely collinear terms, we note that only the leading collinear terms are included, and that their 
coefficient is simply the negative of the leading coefficient of the soft terms.

In the right plot of Fig. \ref{bbhw} we display the total cross section as a function of the charged Higgs mass for various LHC energies. We use MMHT2014 pdf \cite{MMHT2014}. For simplicity we set the value of $\tan\beta$, the ratio of the vevs of the Higgs doublets, equal to unity. The total cross section depends strongly on the charged Higgs mass; it varies over four orders of magnitude in the mass range from 200 GeV to 1000 GeV.

The inset in the right plot of Fig. \ref{bbhw} shows the aNNLO/LO ratio, from which it is clear that the corrections can have a large impact on the cross section. This is in line with significant corrections for related processes, including $tH^-$ production \cite{NKtH}, $tW$ production \cite{tW}, and large-$p_T$ $W$ production \cite {NKW}.

\section{Charged Higgs differential distributions}

\begin{figure}[htb]
\begin{center}
\epsfig{file=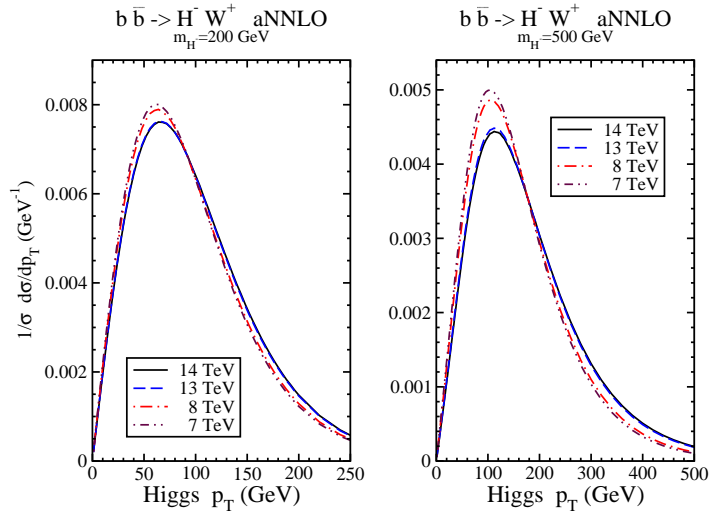,height=3.0in}
\caption{Charged Higgs aNNLO normalized $p_T$ distributions in $H^-W^+$ production.}
\label{ptH}
\end{center}
\end{figure}

Since our resummation formalism provides results for the double-differential cross section, it is feasible to calculate not only total production cross sections, as presented in the previous section, but also differential distributions. In particular, it is of interest to know the transverse momentum and rapidity distributions of the charged Higgs boson.

In Fig. \ref{ptH} we plot the normalized transverse-momentum distributions,
$(1/\sigma) d\sigma/dp_T$, of the charged Higgs boson in $H^-W^+$ production. 
The plot on the left uses 200 GeV for the charged Higgs mass while the plot 
on the right uses 500 GeV. The distributions obviously depend strongly on
the choice of charged Higgs mass. We also observe that the shapes of the 
distributions shift as we move to higher LHC energies. 

\begin{figure}[htb]
\begin{center}
\epsfig{file=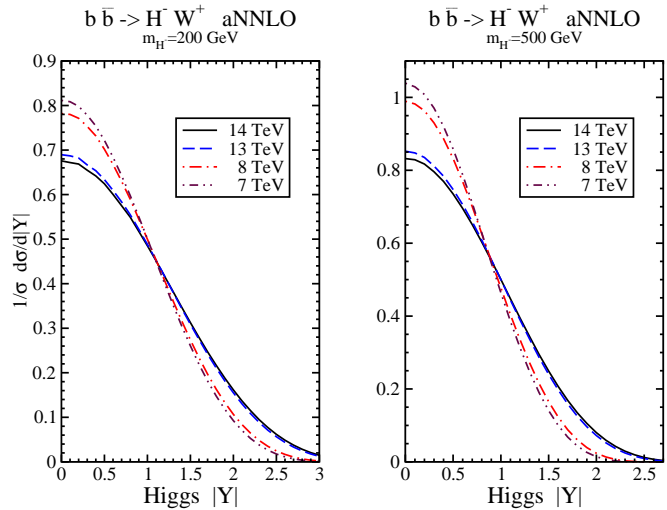,height=3.0in}
\caption{Charged Higgs aNNLO normalized rapidity distributions in $H^-W^+$ production.}
\label{yH}
\end{center}
\end{figure}

In Fig. \ref{yH} we plot the normalized rapidity distributions,
$(1/\sigma) d\sigma/d|Y|$, of the charged Higgs boson. Again, the plot on the 
left is for a 200 GeV charged Higgs mass while the plot on the right is for 
500 GeV. Again, there is a clear dependence of the distributions on charged 
Higgs mass and on LHC energy.

\section*{Acknowledgements}
This material is based upon work supported by the National Science Foundation 
under Grant No. PHY 1519606.

\end{document}